\documentclass [prb,aps,onecolumn,floatfix,showpacs]{revtex4}

\usepackage {amsmath,amssymb,epsfig,color}

\begin{document}

\title
{Transition state method and Wannier functions}

\author{V.I.~Anisimov, A.V.~Kozhevnikov}

\affiliation{Institute of Metal Physics, Russian Academy of
Sciences-Ural Division, 620219 Yekaterinburg GSP-170, Russia}


\date{\today}

\begin{abstract}
We propose a computational scheme for materials where standard
Local Density Approximation (LDA) fails to produce a satisfactory
description of excitation energies. The method uses Slater's
"transition state" approximation and Wannier functions basis set.
We define a correction to LDA functional in such a way that its
variation produces one-electron energies for Wannier functions
equal to the energies obtained in "transition state" constrained
LDA calculations. In the result eigenvalues of the proposed
functional could be interpreted as excitation energies of the
system under consideration. The method was applied to MgO, Si, NiO
and BaBiO$_3$ and gave an improved agreement with experimental
data of energy gap values comparing with LDA.

\end{abstract}
\maketitle
\section{Introduction}
\label{intro}

The one-electron eigenvalues in LDA calculations are Kohn-Sham
energies that were defined as auxiliary Lagrange multipliers in
the problem of Density Functional minimization\cite{Kohn-Sham}. As
such they formally can not be used to calculate spectral
properties of the system because they do not directly correspond
to excitation energies. However for the cases where electron
states are itinerant, for example wide band metals, excitation
spectra calculated with LDA eigenvalues were found to be in good
agreement with experimental photoemission and optical spectra. The
agreement is much worse for band insulators and semiconductors
where LDA gives systematically underestimated values of the energy
gap\cite{jones}. For Mott insulators, for example transition-metal
oxides, LDA calculated spectra could be qualitatively wrong,
giving metallic state while experimentally those systems are wide
gap insulators\cite{cuprates}.

There were many attempts to cure this deficiency of LDA. Among the
most widely used methodes one can mention GW\cite{GW}, SIC\cite{SIC} and
LDA+U\cite{ani91,ani97}. While those approaches have their
advantages there is still no universally accepted calculation
scheme which would be as simple and practical as standard LDA and
a search of better methods continues in scientific community.

The basic problem of a Density Functional Theory (DFT)(and
consequently of the LDA as one of it's approximations) is that DFT
was designed to reproduce  a ground state properties of a
system. In order to obtain excitation energy in the spirit of DFT
one must calculate total energy of the system in the excited
state. Then its difference from the ground state total energy
corresponds to excitation energy. The excited state can be
calculated minimizing LDA functional with the constraining
condition that occupancy of specific one-electron state differs
from its occupancy in the ground state. Such "constrained LDA"
calculations indeed could give good results for excitation
energies\cite{jones}.

Slater had shown \cite{slater} that a good approximation for
excitation energy can be one-electron energy for the specific
one-electron state calculated in a so called "transition state",
where its occupancy is equal 0.5 being half way between the values
for the final and initial state of the excitation process.
Transition state calculations indeed gave excitation energies for
free atoms and ions in a good agreement with experimental
data\cite{slater}.

However this approach can not be directly applied to extended
systems such as solids with translation symmetry where
one-electron states are Bloch functions extended over infinite
crystals. The change of their occupancies produces infinitesimal
change of electron density and so constrained LDA calculations for
"transition state" give exactly the same one-electron energy
values as in a ground state.

A full set of Bloch functions can be replaced by the equivalent
set of Wannier functions produced by some unitary transformation
(see Eqs. (\ref{WF_psi_def}) and (\ref{psi_def})). In contrast to
Bloch states Wannier functions are localized in a specific space
area. If one choose Wannier functions as one-electron states in
"transition state" constrained LDA calculations then there will be
a significant change in the charge density comparing with a ground
state and hence a corresponding correction to the excitation
energy.

The idea of our "generalized transition state method" developed in
the present work is to use "transition state" approximation to
define a modified functional in such a way that Wannier functions
one-electron energies calculated with this functional would have a
meaning of excitation energies for electrons on the corresponding
Wannier function states. This is achieved by adding to LDA
functional a correction changing Wannier functions energies from
LDA values to those obtained in the "transition state"
calculations.

The main effect of "transition state" calculations comparing with
standard LDA is lower one-electron energy values for the occupied
states and higher values for the unoccupied states due to the
decreased occupancy of the former and increased one for the
latter. Then application of the correction should enhance  energy
separation between occupied and empty bands and hence give an
increased value for the energy gap thus improving agreement with
the experimental data.

To test our method we chose a band insulator MgO, semiconductor
Si, Mott insulator NiO and Peierls insulator BaBiO$_3$. In all
those systems LDA gave strongly underestimated energy gap values.
Calculations by "generalized transition state method" resulted in
band structure for all those materials being in a good agreement
with experimental spectroscopy data.

The paper is structured as follows.  In Sec.~\ref{method} we give
the details of our calculation scheme. In Sec.~\ref{results} the
results for the electronic structure of
 MgO, Si, NiO and BaBiO$_3$ obtained by the method developed in this
work are presented and compared with the standard LDA calculations
and experimental data. Finally in Sec.~\ref{conclusion} we close
this work with a conclusion.

\section{Method}
\label{method}

The orbital projection calculation scheme for Wannier functions
(WF's) used in the present work was described in details in the
earlier paper \cite{WF-DMFT} where the LDA+DMFT (DMFT - Dynamical
Mean-Field Theory) method in Wannier function basis set was
proposed. Below we present the main formulas of this scheme.

\subsection{Definition and construction of Wannier functions}
\label{W_def}

The concept of WF's has a very important place in the electron
theory in solids since its first introduction in 1937 by Wannier
.\cite{wannier}  WF's are the Fourier transformation of Bloch
states $|\psi_{i\bf k}\rangle$
\begin{eqnarray}
\label{WF_psi_def} |W_{i}^{\bf T}\rangle& = &
\frac{1}{\sqrt{N}}\sum_{\bf k} e^{-i{\bf kT}}|\psi_{i{\bf
k}}\rangle,
\end{eqnarray}
where $N$ is the number of discrete ${\bf k}$ points in the first
Brillouin zone (or, the number of cells in the crystal) and {\bf
T} is lattice translation vector.

Wannier functions are not uniquely defined for a many-band case
because for a certain set of bands any orthogonal linear
combination of Bloch functions $|\psi_{i\bf k}\rangle$ can be used
in Eq. (\ref{WF_psi_def}). In general it means that the freedom of
choice of Wannier functions corresponds to  freedom of choice of a
unitary transformation matrix $U^{({\bf k})}_{ji}$ for Bloch
functions: \cite{vanderbildt}
\begin{eqnarray}
\label{psi_def} |\widetilde\psi_{i\bf k}\rangle & = & \sum_j
U^{({\bf k})}_{ji} |\psi_{j\bf k}\rangle.
\end{eqnarray}
There is no rigorous way to define $U^{({\bf k})}_{ji}$. This
calls for an additional restriction on the properties of WF's.
Among others Marzari and Vanderbilt \cite{vanderbildt} used the
condition of maximum localization for WF's, resulting in a
variational procedure to calculate $U^{({\bf k})}_{ji}$. To get a
good initial guess the authors of \cite{vanderbildt} proposed
choosing a set of localized trial orbitals $|\phi_n\rangle$ and
projecting them onto the Bloch functions $|\psi_{i\bf k}\rangle$.
It was found that this starting guess is usually quite good. This
fact later led to the simplified calculating scheme in
\cite{pickett} where the variational procedure was abandoned and
the result of the aforementioned projection was considered as the
final step.

For the projection procedure used in the present work one needs to
identify the set of bands and corresponding set of localized trial
orbitals $|\phi_n\rangle$. The choice of bands and
orbitals is determined by the physics of the system under
consideration and will be discussed later.

The set of bands can be defined either by the band indices of the
corresponding Bloch functions ($N_1,...,N_2$), or by choosing the
energy interval ($E_1,E_2$) in which the bands are located.
Non-orthogonalized WF's in reciprocal space $|\widetilde{W}_{n\bf
k}\rangle$ are then projection of the set of site-centered
atomiclike trial orbitals $|\phi_n\rangle$ on Bloch functions
$|\psi_{i\bf k}\rangle$ of the chosen bands defined by band
indices ($N_1$ to $N_2$) or by energy interval ($E_1,E_2$):
\begin{eqnarray}
\label{WF_psi} |\widetilde{W}_{n\bf k}\rangle & \equiv &
\sum_{i=N_1}^{N_2} |\psi_{i\bf k}\rangle\langle\psi_{i\bf
k}|\phi_n\rangle = \sum_{i(E_1\le \varepsilon_{i}({\bf k})\le
E_2)} |\psi_{i\bf k}\rangle\langle\psi_{i\bf k}|\phi_n\rangle.
\end{eqnarray}

In the present work we have used LMTO method \cite{LMTO} to solve
a band structure problem and the trial orbitals $|\phi_n\rangle$
were LMTO's. Note that in the multi-band case a WF in reciprocal
space $|\widetilde{W}_{n\bf k}\rangle$ does not coincide with the
eigenfunction $|\psi_{n\bf k}\rangle$ due to the summation over
band index $i$ in Eq. (\ref{WF_psi}). One can consider them as
Bloch sums of WF's analogous to the basis function Bloch sums
$\phi_j^{\bf k}({\bf r})$ (Eq. (\ref{psik})). The coefficients
$\langle\psi_{i\bf k}|\phi_n\rangle$ in Eq. (\ref{WF_psi}) define
(after orthonormalization) the unitary transformation matrix
$U^{({\bf k})}_{ji}$ in Eq. (\ref{psi_def}).

In any DFT method the Kohn-Sham orbitals are expanded through
the certain basis:
\begin{eqnarray}\label{psi}
|\psi_{i\bf k}\rangle & = & \sum_{\mu} c^{\bf k}_{\mu
i}|\phi_{\mu}^{\bf k}\rangle.
\end{eqnarray}
The basis functions of the LMTO method are Bloch sums of the cite
centered orbitals:
\begin{eqnarray} \label{psik}
\phi_{\mu}^{\bf k}({\bf r}) & = & \frac{1}{\sqrt{N}}
\sum_{\bf T} e^{i\bf kT} \phi_{\mu}({\bf r}-{\bf R}_q-{\bf T}),
\end{eqnarray}
where $\mu$ is the combined index representing $qlm$ ($q$ is the
atomic number in the unit cell, $lm$ are orbital and magnetic
quantum numbers), ${\bf R}_q$ is the position of atom in the
unit cell.

For the {\it orthogonal} LMTO basis $c^{\bf k}_{\mu
i}=\langle\phi_{\mu}|\psi_{i \bf k}\rangle $ and hence

\begin{eqnarray}
\label{WF} |\widetilde{W}_{n\bf k}\rangle & = &
\sum_{i=N_1}^{N_2} |\psi_{i\bf k}\rangle c_{ni}^{{\bf k}*}
 =  \sum_{i=N_1}^{N_2} \sum_{\mu} c^{\bf k}_{\mu i} c_{ni}^{{\bf k}*}
|\phi_{\mu}^{\bf k}\rangle  = \sum_{\mu} \tilde{b}^{\bf k}_{\mu n}
|\phi_{\mu}^{\bf k}\rangle,
\end{eqnarray}
with
\begin{equation}
\label{coef_b} \tilde{b}^{\bf k}_{\mu n} \equiv \sum_{i=N_1}^{N_2}
c^{\bf k}_{\mu i} c_{ni}^{{\bf k}*}.
\end{equation}
For a nonorthogonal basis set orthogonalization of the Hamiltonian
must be done before using Eq.(\ref{coef_b}).

In order to orthonormalize the WF's Eq. (\ref{WF}) one needs to
calculate the overlap matrix $O_{nn'}({\bf k})$
\begin{eqnarray}
\label{O-S} O_{nn'}({\bf k})&\equiv& \langle\widetilde{W}_{n\bf
k}|\widetilde{W}_{n'\bf k}\rangle = \sum_{i=N_1}^{N_2} c^{\bf
k}_{ni} c_{n'i}^{{\bf k}*},
\end{eqnarray}
and its inverse square root $S_{nn'}({\bf k})$
\begin{eqnarray}
\label{SS} S_{nn'}({\bf k}) &\equiv& O^{-1/2}_{nn'}({\bf k}).
\end{eqnarray}
In the derivation of Eq. (\ref{O-S}) the orthogonality of Bloch states
$\langle\psi_{n\bf k}|\psi_{n'\bf k} \rangle=\delta_{nn'}$ was
used.

From Eqs. (\ref{WF}) and (\ref{SS}) the orthonormalized WF's in
${\bf k}$ space $|W_{n\bf k}\rangle$ can be obtained as
\begin{eqnarray}
\label{WF_orth} |W_{n\bf k}\rangle = \sum_{n'} S_{nn'}({\bf k})
|\widetilde{W}_{n'\bf k}\rangle =\sum_{i=N_1}^{N_2} |\psi_{i\bf
k}\rangle \bar{c}_{ni}^{{\bf k}*}  = \sum_{\mu} b^{\bf k}_{\mu n}
|\phi_{\mu} ^{\bf k}\rangle,
\end{eqnarray}
with
\begin{eqnarray}
\label{coef_WF_Bloch} \bar{c}_{ni}^{{\bf k}*}\equiv
\langle\psi_{i\bf k}|W_{n\bf k}\rangle =\sum_{n'} S_{nn'}({\bf k})
c_{n'i}^{{\bf k}*},
\end{eqnarray}
\begin{eqnarray}
\label{coef_WF_LMTO} b^{\bf k}_{\mu n} \equiv
\langle\phi_{\mu}^{\bf k}|W_{n\bf k}\rangle= \sum_{i=N_1}^{N_2}
c^{\bf k}_{\mu i} \bar{c}_{ni}^{{\bf k}*}.
\end{eqnarray}

The real space site-centered WF's at the origin
$|W_{n}^{\bf0}\rangle$ are given by the Fourier transform of
$|W_{n\bf k}\rangle$ (Eq. (\ref{WF_psi_def})) with ${\bf T}=0$.
From Eqs. (\ref{WF_orth}) and (\ref{psik}) one finds
\begin{eqnarray}
\label{real_WF}  W_n({\bf r})= \frac{1}{\sqrt{N}}\sum_{\bf k}
\langle{\bf r}|W_{n\bf k}\rangle
 & = & \sum_{{\bf T},\mu}\biggl(\frac{1}{N}\sum_{\bf k}
 e^{i\bf kT} b_{\mu n}^{\bf k}\biggr)
\phi_{\mu}({\bf r}-{\bf T}) \nonumber \\
 & = &  \sum_{{\bf T},\mu} w'(n,\mu,{\bf T}) \phi_{\mu}({\bf r}-{\bf T})\\\nonumber
 & = &\sum_{s} w(n,s) \phi_{\alpha(s)}({\bf r}-{\bf T}_s), \\ \nonumber
\end{eqnarray}
where $w'$ and $w$ are the expansion coefficients of WF in terms
of the corresponding LMTO orbitals, in particular,
\begin{eqnarray}
w(n,s)  = \frac{1}{N}\sum_{\bf k} e^{i{\bf kT}_s}
b_{\alpha(s)n}^{\bf k}.
\end{eqnarray}
Here $s$ is an index counting the orbitals of the neighboring
cluster for the atom where orbital $n$ is centered (${\bf T}_s$ is
the corresponding translation vector, $\alpha(s)$ is a combined
$qlm$ index). The explicit form of the real space WF
Eq. (\ref{real_WF}) can be used to produce, e.g., shapes of chemical
bonds. For other applications only the matrix elements of the
various operators in the basis of WF Eq. (\ref{WF_orth}) are needed.

Using Eqs. (\ref{WF_orth}), (\ref{coef_WF_Bloch}), and (\ref{coef_WF_LMTO})
one can find energies of WF's:

\begin{eqnarray} \label{E_WF}
E^{WF}_{n} & = & \langle W_{n}^{\bf T}|
  \biggl(\sum_{i,\bf k} |\psi_{i\bf k}\rangle
    \epsilon_{i}({\bf k})\langle\psi_{i\bf k}|
  \biggr)|W_{n}^{\bf T}\rangle  \nonumber \\
           & = & \frac{1}{N}
       \sum_{\bf k}\sum_{i=N_1}^{N_2} \bar{c}^{\bf k}_{ni}
      \bar{c}_{n'i}^{{\bf k}*}\epsilon_{i}({\bf k})
\end{eqnarray}
and their occupancies:
\begin{eqnarray} \label{Q_WF}
Q^{WF}_{n} & = & \langle W_{n}^{\bf T}|
  \biggl(\sum_{i,\bf k} |\psi_{i\bf k}\rangle
    \theta(E_f - \epsilon_{i}({\bf k}) )\langle\psi_{i\bf k}|
  \biggr)|W_{n}^{\bf T}\rangle  \nonumber \\
           & = & \frac{1}{N}
       \sum_{\bf k}\sum_{i=N_1}^{N_2} \bar{c}^{\bf k}_{ni}
      \bar{c}_{n'i}^{{\bf k}*}\theta(E_f - \epsilon_{i}({\bf k}) ),
\end{eqnarray}
where $\epsilon_{i}({\bf k})$ is the eigenvalue for a particular band,
$\theta(x)$ is the step function, $E_f$ is the Fermi energy.

The transformation from LMTO to WF basis set is defined by the
explicit form of WF's Eqs. (\ref{WF_orth}), (\ref{coef_WF_LMTO}),
and by the expressions for matrix elements of the Hamiltonian and
density matrix operators in WF basis (Eqs. (\ref{E_WF}) and
(\ref{Q_WF})). The transformation from WF  to LMTO basis can also
be defined using Eq. (\ref{WF_orth}). Such transformation is
needed in calculations using correction potential in the form of
Eq. (\ref{H_corr}) and for constrained LDA calculations
determining "transition state" energies of WF's. For example if
constrain potential is diagonal in WF basis ($H_{nn'}=\Lambda
_n\delta_{nn'}$), then in LMTO basis its matrix elements can be
calculated via:
\begin{eqnarray} \label{H_constrain}
\widehat{H}_{constr}  =  \sum_{n,{\bf T}} |W_{n}^{\bf T} \rangle
\Lambda_n \langle W_{n}^{\bf T}|
\end{eqnarray}
\begin{equation} \label{W-b}
|W_{n}^{\bf T} \rangle  =  \sum_{j,{\bf k}} e^{-i{\bf kT}} b^{\bf
k}_{jn} |\phi_j^{\bf k}\rangle
\end{equation}
\begin{eqnarray} \label{H_constrain2}
H_{\mu\nu}(\bf k) & = & \langle \phi_{\mu}^{\bf k}
|\widehat{H}_{constr}| \phi_{\nu}^{\bf k}\rangle\\ \nonumber & = &
\sum_n b_{\mu n}^{\bf k} \Lambda_n b_{\nu n}^{\bf k*}
\end{eqnarray}

\subsection{"Generalized transition state" method}
\label{GTS}

Transition state calculation scheme proposed by Slater
\cite{slater} allows to calculate excitation energy for the
process of adding (removing) an electron to (from) the system from (to)
the infinity
where potential is supposed to be equal zero. For that one should
calculate LDA eigenvalue (Kohn-Sham equations eigenvalue) of the
corresponding one-electron state with its occupancy equal 0.5. In
other words for the occupied states the occupancy is reduced by
one half and for the empty ones it is increased by one half.

This scheme can be derived in the following way. The total energy
difference between final and initial states for the process of
electron addition to the one-electron state $j$ can be calculated as an
integral of total energy derivative over occupancy $q_j$ (this
derivative is\cite{jones} the corresponding LDA eigenvalue
$\epsilon_j=\frac{\partial E}{\partial q_j}$):
\begin{eqnarray}
\label{trans} E(q_j=1)-E(q_j=0) & = &  \int\limits_0^1 dq_j \left(
\frac{\partial E}{\partial q_j}\right) =\\ \nonumber
 =\int\limits_0^1 dq_j  \epsilon_{j}(q_j)  \approx
 \epsilon_{j}(0.5)
\end{eqnarray}
It means that the {\em one-electron} energy (LDA functional
eigenvalue) calculated with its occupancy value half way between
final and initial states is a good approximation for the {\em
total} energy difference (excitation energy). This equation
becomes exact if LDA eigenvalue $\epsilon_{j}(q_j)$ is a linear
function of occupancy $q_j$ ($\frac{\partial
\epsilon_{j}(q_j)}{\partial q_j}=const$), which is usually with a
good accuracy true. In this case for empty state ($q_j=0$):
\begin{eqnarray}
\label{const}
\epsilon_{j}(0.5)& = &
\epsilon_{j}(0)+\frac{1}{2}\frac{\partial
\epsilon_{j}(q_j)}{\partial q_j}
\end{eqnarray}
For occupied states ($q_j=1$) a sign plus in Eq. (\ref{const})
will be replaced by minus. The general formula will be:
\begin{eqnarray}
\label{const2} \epsilon_{j}(0.5)& = &
\epsilon_{j}(q_j)+(\frac{1}{2}-q_j)\frac{\partial
\epsilon_{j}(q_j)}{\partial q_j}
\end{eqnarray}
 So effect of "transition state" correction to LDA values is to
increase energy for empty states (addition energy) and to decrease
it for the occupied states (removal energy). That will also
results in a larger value of energy for the excitation from
occupied to empty states of the system.

Transition state method can be reformulated in a functional
formalism. For that equation (\ref{const2}) should be obtained via
variation of an auxiliary functional. This "transition state"
functional $E_{TS}$ is defined by adding to LDA functional
$E_{LDA}[\rho]$ the correction term\cite{meaning} depending on the
occupancies $q_j$:
\begin{eqnarray}
\label{functional-trans} E_{TS}= E_{LDA}[\rho]-\frac{1}{2} \sum_j
\frac{\partial \epsilon_{j}(q_j)}{\partial q_j}(q_j-\frac{1}{2})^2
\end{eqnarray}

All eigenvalues obtained from this functional will automatically
have a correction $(\frac{1}{2}-q_j)\frac{\partial
\epsilon_{j}(q_j)}{\partial q_j}$ to LDA eigenvalues like in the
right part of Eq. (\ref{const2}). Then one calculation for the
{\it{ground state}} of the functional $E_{TS}$ gives result
equivalent to a set of "transition state" calculations for every
one-electron state. However one still needs to run a set of
constrained LDA calculations in order to determine derivatives
$\frac{\partial \epsilon_{j}(q_j)}{\partial q_j}$.

 The "transition state" method to
calculate excitation energies has proved to be sufficiently
successful for small size systems like atoms and ions\cite{slater}
but it can not be directly applied to solids. Bloch functions are
extended over the crystal and a change of their occupancy will
have a negligible effect on the charge density. In this case
"transition state" calculations will give the same one-electron
energies as the ground state results.

For a case of fully occupied bands transformation from a set of
Bloch states to Wannier functions does not change charge density
distribution. It is more than that, many-electron function defined
as a Slater determinant constructed from one-electron Bloch
functions is identical to the Slater determinant made out of
Wannier functions. (The determinant value is not changed in the
result of adding to one of its rows a linear combination of other
rows and Wannier functions are by definition linear combinations of
Bloch functions (see Eqs. (\ref{WF_psi_def}) and
(\ref{psi_def}))). Then ground state properties are invariant to
such transformation.

However for excited states there is an important difference
between Bloch and WF representations. As WF is localized, a change
of its occupancy will result in a finite charge density
modification in the area of its localization and hence "transition
state" calculations will result in a significant correction for
the WF one-electron energy. While in standard LDA one-electron
energies are Kohn-Sham energies that were defined as auxiliary
Lagrange multipliers in the problem of minimizing Density
Functional\cite{Kohn-Sham}, WF energies obtained in "transition
state" calculations have a meaning of excitation energies for an
electron on the corresponding WF states.


The idea of our "generalized transition state" method (GTS) is to
replace in the original formulation of "transition state" method
(Eqs. \ref{trans}-\ref{const2})) a set of one-electron states in a
form of Bloch functions for an infinite crystal by an equivalent
set of Wannier functions. The corresponding functional will be
analogous to Eq. (\ref{functional-trans}) but with occupancies
$Q^{WF}_{n}$ (Eq. (\ref{Q_WF})) and energies $E^{WF}_{n}$ (Eq.
(\ref{E_WF})) corresponding to Wannier functions:
\begin{eqnarray}
\label{functional-GTS} E_{GTS}= E_{LDA}-\frac{1}{2} \sum_n
\frac{\partial E^{WF}_{n}}{\partial
Q^{WF}_{n}}(Q^{WF}_{n}-\frac{1}{2})^2
\end{eqnarray}

The variation of the correction term in (\ref{functional-GTS})
will produce a correction Hamiltonian $\widehat{H}_{corr}$ in the
form of projection operator:
\begin{eqnarray}
\label{H_corr}
\widehat{H}_{corr} & = & \sum_{n\bf T} |W_{n}^{\bf
T} \rangle \delta V_{n} \langle W_{n}^{\bf T}|.
\end{eqnarray}
Then Hamiltonian operator for "generalized transition state"
method is given by:
\begin{eqnarray}
\label{H_corr2} \widehat{H}_{GTS} & = &
\widehat{H}_{LDA}+\widehat{H}_{corr}.
\end{eqnarray}

$|W_{n}^{\bf T} \rangle$ in Eq. (\ref{H_corr}) are Wannier
functions (Eq. (\ref{W-b})) and $\delta V_{n}$ are defined as a
difference between Wannier functions one-electron energies
calculated in "transition state" and ground state (see Eq.
\ref{const2}):
\begin{eqnarray}
\label{dV-GTS} \delta V_{n} = \frac{\partial E^{WF}_{n}}{\partial
Q^{WF}_{n}}(\frac{1}{2}-Q^{WF}_{n})
\end{eqnarray}
The values of derivatives $ \frac{\partial E^{WF}_{n}}{\partial
Q^{WF}_{n}}$ (or equivalently $\delta V_{n}$ themselves) should be
determined in constrained LDA calculations.

The calculation scheme is the following (see Fig.\ref{wf_scheme}).
To define correction operator (Eq. (\ref{H_corr})) one needs to
know a set of "transition state" corrections to WF's energy values
$\delta V_n$ (Eq. (\ref{dV-GTS})) and explicit form of WF's
determined by expansion coefficients in the basis orbitals (LMTO)
$b^{\bf k}_{\mu n}$ (Eq. (\ref{coef_WF_LMTO})). Both $\delta V_n$
and $b^{\bf k}_{\mu n}$ should be calculated self-consistently. On
every self-consistency loop Bloch functions $|\psi_{i\bf
k}\rangle$ calculated with "generalized transition state" method
Hamiltonian $\widehat{H}_{GTS}$ (Eq. (\ref{H_corr2})) are used to
define new Wannier functions $|W_{n\bf k}\rangle$ via Eq.
(\ref{WF}) to get a new set of coefficients $b^{\bf k}_{\mu n}$.
Then a series of constrained LDA calculations (using constrain
potential in the form of Eq. (\ref{H_constrain})) for every type
of WF $|W_{n}^{0} \rangle$ (Eq. (\ref{real_WF})) is performed
where the occupancy (Eq. (\ref{Q_WF})) of this particular WF is
kept to be 0.5. The energy of this WF's is then calculated using
Eq. (\ref{E_WF}). The new $\delta V_n$ is defined as a difference
between this "transition state" value and the corresponding value
from calculation where the WF occupancy is the same as in ground
state.

\begin{widetext}
\onecolumngrid
  \begin{figure}
    \includegraphics[clip=true,width=0.7\textwidth]{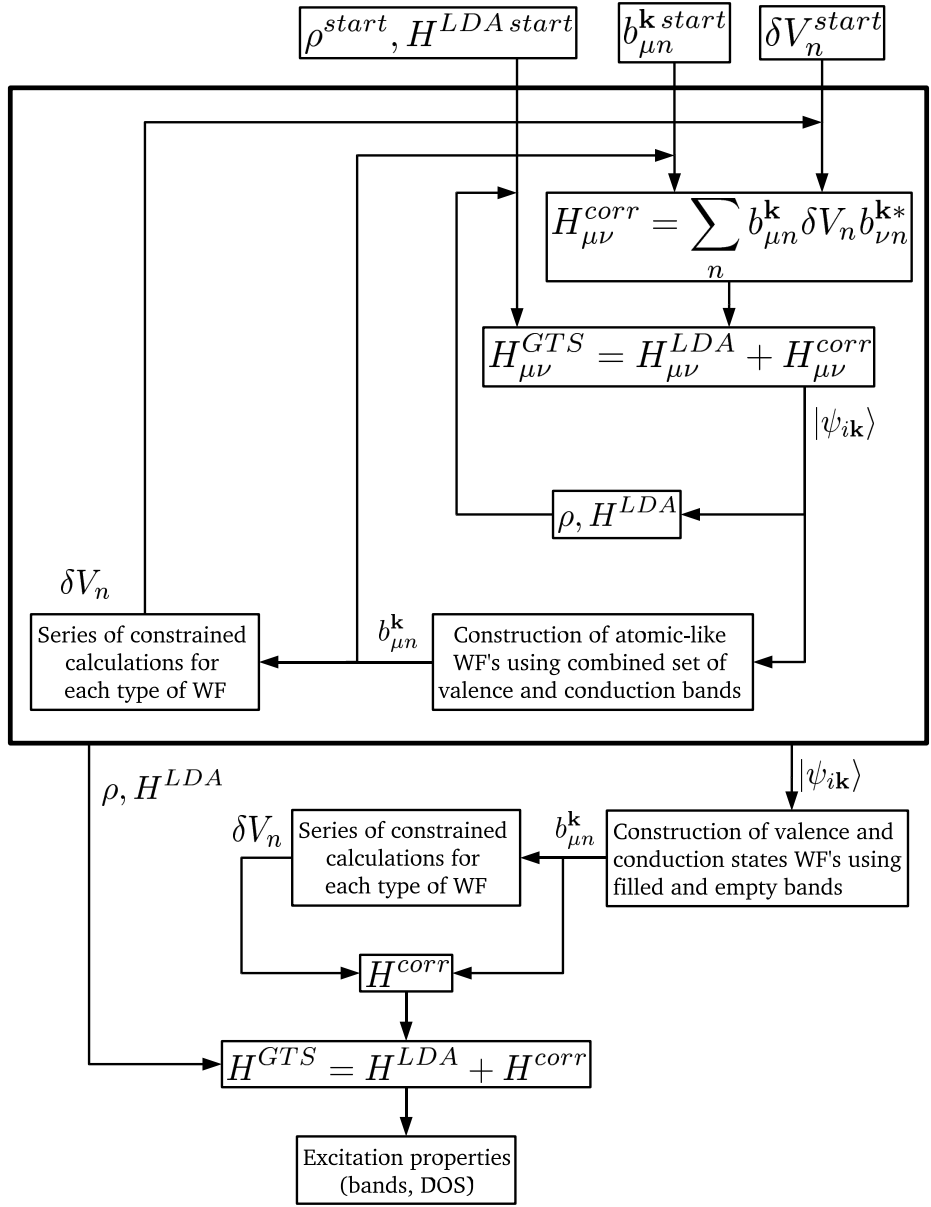}
    \caption{The calculation scheme of "generalized transition state" method.}
    \label{wf_scheme}
  \end{figure}
\end{widetext}

The important question is what set of bands and trial orbitals
should be used in projection procedure (Eq. (\ref{WF_psi})) to
calculate WF's. In band insulators and semiconductors valence band
corresponds to bonding states and conduction band to antibonding
ones. So if one will calculate WF's via projection procedure
separately for valence and conduction bands (summation over band
indices in Eq. (\ref{WF_psi}) is running over occupied bands only
for valence WF's and over empty bands for conduction states WF's)
then the results will be bonding and antibonding functions
extending over neighboring atoms. But if WF's will be obtained
using a full set of valence and conduction bands (summation over
band indices in Eq. (\ref{WF_psi}) is running over combined set of
occupied and empty  bands both for valence and conduction states
WF's), then the opposite sign contributions on neighboring atoms
from bonding and antibonding functions cancel each other and the
resulting WF resembles an original atomic orbital.

For spectral properties where excitation occurs from (to) valence
(conduction) states WF should represent the corresponding bonding
(antibonding) functions. Then the most natural choice would be to
use in projection procedure (Eq. (\ref{WF_psi})) two separate sets
of occupied and empty bands and two sets of atomic-like orbitals
that give a strongest contribution to the corresponding bands. For
example in MgO case the orbitals for occupied bands would be
oxygen 2p and for empty bands Mg 3s and 3p orbitals. The
application of the correction operator (Eq. (\ref{H_corr})) will
shift relative energies of the bands but will not change
significantly wave functions themselves. They will remain the same
as were obtained in LDA calculations. Then charge and spin
electron density will be not modified by GTS correction.

This approximation could be satisfactory for semiconductors and
band insulators where the only problem of LDA is underestimation
of excitation energies. However for Mott insulators LDA
calculations could give even ground state properties qualitatively
wrong. For example for parent high-T$_c$ compound La$_2$CuO$_4$
LDA gives nonmagnetic metallic solution while experimentally it is
an antiferromagnetic insulator\cite{cuprates}. For NiO LDA gave
strongly underestimated value of magnetic moment on Ni ion (1.0
$\mu_B$ versus experimental value $\approx$1.8-1.9 $\mu_B$
\cite{M-NiO}). The problem with these systems could be traced to
underestimation in LDA of the energy of {\it{ virtual}}
excitations from occupied oxygen 2p to empty transition-metal 3d
orbitals. This leads to overestimation of hybridization between
those orbitals and hence decreases a tendency to magnetic moment
formation.

In order to correct the LDA underestimation of {\it{ virtual}}
excitations energies one must use in "transition state"
calculations WF's resembling pure atomic orbitals such as oxygen 2p
and transition-metal 3d orbitals and not their bonding
(antibonding) combinations which are valence (conduction state)
WF's. As it was explained above for that one should choose in
projection procedure (Eq. (\ref{WF_psi})) the combined single set
of valence and conduction bands instead of the two separate sets
of occupied and empty bands. The "transition state" calculations
for these Wannier functions will give the result which have a
meaning of energies for {\it{ virtual}} excitations between atomic
orbital states and not real excitations which happen in
spectroscopy experiments from (to) valence (conduction) band
states. The corresponding correction in the form of Eq.
(\ref{H_corr}) will increase the energy of {\it{ virtual}}
excitations from occupied oxygen 2p to empty transition-metal 3d
orbitals and hence decrease hybridization between those orbitals
that should enhance a tendency to magnetic moment formation.

That means that a different definition of WF's is needed in
calculations for ground state properties (atomic orbital WF's
obtained by projection on the combined single set of valence and
conduction bands) and in calculations for spectral properties
(valence (conduction state) WF's projected by using separately
valence bands for occupied states and conduction bands for
unoccupied ones).


In the present work we performed two steps of calculation for
every system (see Fig.\ref{wf_scheme}). In the first one (part of
the Fig.\ref{wf_scheme} scheme bounded in a bold line rectangle)
we chose a single combined set of bands to define WF's resembling
pure atomic orbitals. Calculations are repeated till achieving
full self-consistency in charge and spin densities $\rho(\vec r)$
(defining one-electron LDA potential and hence parameters of LDA
Hamiltonian $H_{LDA}$), set of "transition state" corrections to
WF's energies $\delta V_n$ and WF's themselves determined by
expansion coefficients $b^{\bf k}_{\mu n}$ (Eq.
(\ref{coef_WF_LMTO})). In the result ground state properties of
the system are calculated which are modified from those in LDA
solution via "transition state" correction to energy parameters of
{\it{ virtual}} excitations. The next step is to calculate
"transition state" correction to the energies of {\it{real}}
excitation observed in spectroscopic measurements. At this stage
we fix obtained on the first step Bloch functions $|\psi_{i\bf
k}\rangle$ and parameters of LDA Hamiltonian $H_{LDA}$. From these
Bloch functions $|\psi_{i\bf k}\rangle$ new valence and conduction
state Wannier functions are calculated (new expansion coefficients
$b^{\bf k}_{\mu n}$ (Eq. (\ref{coef_WF_LMTO}))) in projection
procedure (Eq. (\ref{WF_psi})) using two separate sets of occupied
and empty bands.  These new WF's define a new set of "transition
state" corrections to WF's energies $\delta V_n$ via constrained
LDA transition state calculations and hence a new correction
Hamiltonian (Eq. (\ref{H_corr})) which is used to calculate
spectral properties of the system.

The above described calculation scheme with two different
definitions of Wannier functions could seem to be overcomplicated,
but we have found that no single set of WF's can describe both
ground state and excitation properties. In general, the choice of
procedure defining Wannier functions is dictated by the physics
which we want to describe in our calculations.

\section{Results and discussion}
\label{results}
 In order to test proposed above "generalized
transition state" (GTS) method we have performed calculations for
four systems representing various types of electronic structure:
simple metal oxide band insulator MgO, covalent bond semiconductor
Si, transition metal oxide  Mott insulator NiO and Peierls
insulator BaBiO$_3$. In all those cases we have obtained
significant change of electronic structure leading to good
agreement of calculated and experimental energy gap values (see
Table \ref{result}).

\begin{table}

\begin{tabular*}{0.45\textwidth}{l c c c}
\hline \hline
\hspace{1.7cm}  & \quad\, LDA \quad\, & \quad\, GTS \quad\, & \quad\, Expt. \quad\, \\
\hline
MgO       & 5.04  &  7.73  & 7.83\footnote{Ref.~\onlinecite{MgO_expt}}   \\
Si        & 0.44  &  1.04  & 1.17\footnote{Ref.~\onlinecite{Si_expt}}   \\
NiO       & 0.11  &  3.76  & 4.0-4.3\footnote{Ref.~\onlinecite{NiO_expt2}-\onlinecite{NiO_expt3}}   \\
BaBiO$_3$ & 0.15  &  0.51  & 0.48\footnote{Ref.~\onlinecite{BaBiO3_expt}}   \\
\hline \hline
\end{tabular*}
\caption{Comparison of calculated and experimental energy gap
values (eV)} \label{result}
\end{table}

\subsection{MgO}
\begin{figure}
  \includegraphics[clip=true,width=0.45\textwidth]{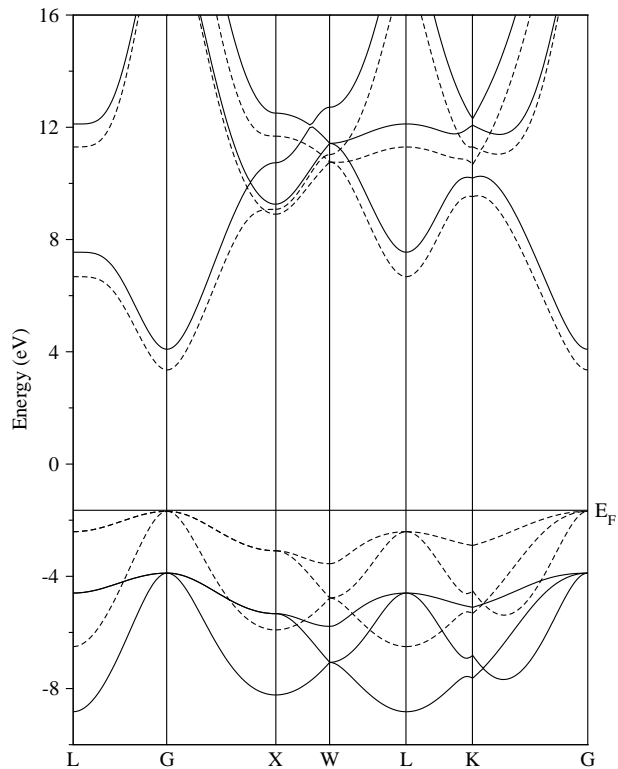}
  \caption{MgO band structure. Dashed lines shows LDA results and solid lines
  correspond to GTS calculations}
  \label{MgO}
\end{figure}

Simple metal oxide MgO is a good example of band insulator. It has
a cubic NaCl crystal structure with lattice parameter equal to
4.21 \AA. Experimental value of energy gap (7.83
eV\cite{MgO_expt}) is nearly 3 eV larger then LDA calculated value
(5.04 eV in our calculations in good agreement with other works
\cite{MgO-LDA}). The choice of energy bands and orbitals needed
for calculation of Wannier functions via Eq.(\ref{WF_psi}) is
straightforward in this case: oxygen 2p orbitals for three
occupied valence bands and magnesium 3s and 3p orbitals for
unoccupied conduction bands. Constrained LDA calculations for
transition state gave the following corrections to the energies of
Wannier functions: $\delta V_{Mg3s} = 1.44$ eV, $\delta V_{Mg3p} =
1.81$ eV, $\delta V_{O2p} = -2.41$ eV . Stronger effect of GTS
correction for O$2p$ states comparing with Mg$3s,3p$ WF's could be
understood taking into account more extended nature of magnesium
orbitals comparing with oxygen ones. On Fig.\ref{MgO} band
structure calculated by GTS method is presented together with the
results of standard LDA calculations. According to the $\delta V$
correction values valence bands formed by O$2p$ states are pushed
down and conduction Mg$3s,3p$ bands are pushed up increasing the
energy gap value to good agreement with experimental data (see
Table \ref{result}). Please note that GTS correction potential
(Eq. (\ref{H_corr})) is not a rigid shift of bands and so the
difference between LDA and GTS gap values is not simply given by
the corresponding difference of $\delta V$ correction values.

\subsection{Si}
\begin{figure}
  \includegraphics[clip=true,width=0.45\textwidth]{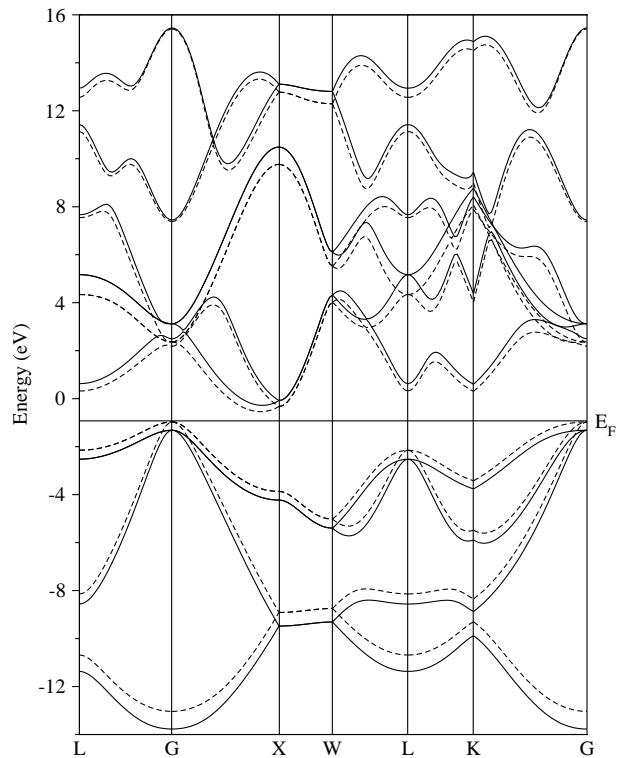}
  \caption{Si band structure. Dashed lines shows LDA results and solid lines
  correspond to GTS calculations}
\label{Si bands}
\end{figure}

 While MgO is an ionic compound, silicon represents
a simplest case of a covalent bond semiconductor. Si has diamond
crystal structure with lattice parameter equal to 5.43 \AA. For
this system we used four types of Wannier functions: $Si3s_1$ and
$Si3p_1$ calculated for the valence bands and $Si3s_2$ and
$Si3p_2$ for conduction bands. The corresponding values of GTS
potential (Eq. (\ref{H_corr})) corrections obtained in constrained
LDA transition state calculations are: $\delta V_{Si3s_1} = -0.82$
eV, $\delta V_{Si3p_1} = -0.44$ eV, $\delta V_{Si3s_2} = 0.28$ eV,
$\delta V_{Si3p_2} = 0.71$ eV. The resulting band structure is
shown on Fig.\ref{Si bands}. Negative potential correction for
valence bands and a positive one for conduction bands produce a
sizable increasing (0.60 eV) of the energy gap value resulting in
a good agreement with experimental data (see Table \ref{result}).

\subsection{NiO}
In contrast to energy gap values the ground state properties for
MgO and Si are satisfactory reproduced by LDA and the first step
of GTS calculations with atomic-orbital-like WF's (see Sec.
\ref{GTS} and Fig.\ref{wf_scheme}) did not lead to significant
changes comparing with pure LDA. However for transition metal
oxides even the type of the ground state can be given wrong by
LDA, like for cuprates\cite{cuprates}. For nickel oxide NiO (cubic
NaCl crystal structure with lattice parameter equal to 4.17 \AA)
the LDA error is not so severe: the LDA solution is an
antiferromagnetic insulator in agreement with experiment. However
LDA  magnetic moment value for Ni ion is strongly
underestimated:1.0 $\mu_B$ versus experimental value
$\approx$1.8-1.9 $\mu_B$ \cite{M-NiO}. In this case the first step
of GTS calculations  responsible for the "transition state"
correction to energy parameters of {\it{ virtual}} excitations was
essential: calculated magnetic moment value was increased in the
results of GTS correction from 1.0 $\mu_B$ to 1.8 $\mu_B$ in good
agreement with experimental data. There were five different types
of Wannier functions in calculations for NiO:
Ni3d-$t_{2g}\uparrow$,Ni3d-$e_{g}\uparrow$,
Ni3d-$t_{2g}\downarrow$, Ni3d-$e_{g}\downarrow$ (the only
unoccupied states for NiO) and O2p (by symmetry O2p states are not
spin-polarized). Constrained LDA calculations for transition state
gave the following corrections to the energies of Wannier
functions:$\delta V_{Ni3d-t_{2g}\uparrow} = -1.97$ eV, $\delta
V_{Ni3d-e_{g}\uparrow} = -1.97$ eV, $\delta
V_{Ni3d-t_{2g}\downarrow} = -2.39$ eV, $\delta
V_{Ni3d-e_{g}\downarrow} = 2.11$ eV and $\delta V_{O2p} = -1.14$
eV.

\begin{figure}
  \includegraphics[clip=true,width=0.45\textwidth]{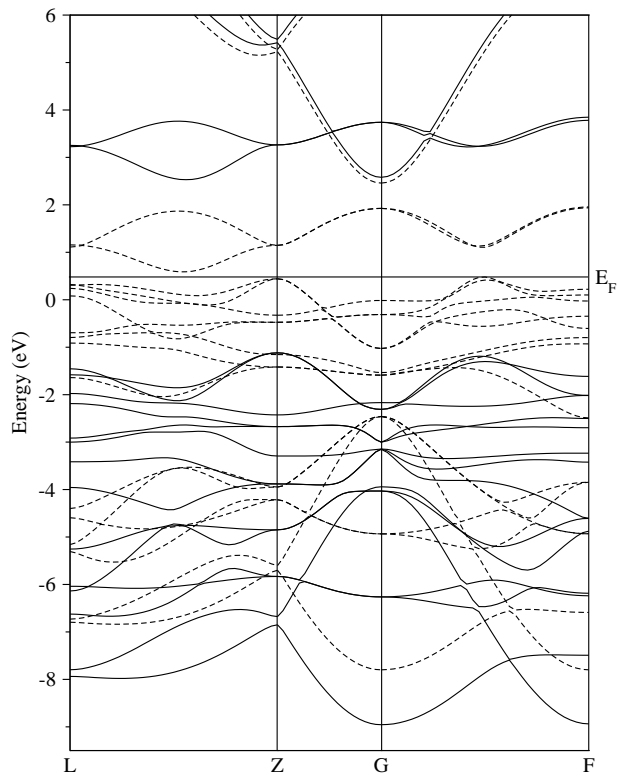}
  \caption{NiO band structure. Dashed lines shows LDA results and solid lines
  correspond to GTS calculations}
  \label{NiO-bands}
\end{figure}

The energy band dispersions obtained in GTS calculations together
with LDA bands are shown on Fig.\ref{NiO-bands}. The energy gap
value was increased dramatically: from tiny 0.11 eV to a 3.76 eV
value in good agreement with experimental data (see Table
\ref{result}). Not only the energies of unoccupied
$Ni3d-e_{g}\downarrow$ states were pushed up and those for
occupied bands pushed down as it was the case for MgO and Si. The
different values of $\delta V$ correction for occupied Ni3d and
O2p states result in a smaller energy separation between the
corresponding bands in the occupied part of the calculated DOS
(see Fig.\ref{NiO-DOS}). Note also decreased strength of admixture
of $Ni3d-e_{g}\downarrow$ states to the oxygen band energy area
resulting in a more ionic nature of GTS solution comparing with
pure LDA and hence magnetic moment value 1.8 $\mu_B$ closer to
pure ionic value 2.0 $\mu_B$

\subsection{BaBiO$_3$}
BaBiO$_3$ is an interesting example of the Peierls insulator or
"negative $U$" system\cite{And}. Formal valency of bismuth is +4
and this corresponds to the half-filled Bi6s shell. Stable valent
states for Bi are +3 and +5 and those chemical arguments are often
used to explain the experimental distorted cubic perovskite
crystal structure of BaBiO$_3$\cite{BaBiO3-cryst}. In addition to
the tilting of BiO$_6$ octahedra there is also so called
"breathing" distortion producing inequivalent Bi1 and Bi2
crystallographic positions with expanded and contracted Bi-O bond
lengths. Bi1 can be associated with Bi$^{+3}$ and Bi2 with
Bi$^{+5}$. This distortion leads to opening an energy gap so
BaBiO$_3$ can be seen as a three-dimensional Peierls insulator.
The Fermi level in LDA calculations for ideal cubic perovskite
crystal structure crosses half-filled band of Bi6s symmetry which
can be described by an effective half-filled Hubbard model. An
experimentally observed instability toward formation sites with
empty (Bi$^{+5}$) and completely filled (Bi$^{+3}$) Bi6s shell can
be interpreted as a negative value for a Coulomb interaction
parameter $U$.

\begin{widetext}
\onecolumngrid
\begin{figure}
  \includegraphics[clip=true,width=0.95\textwidth]{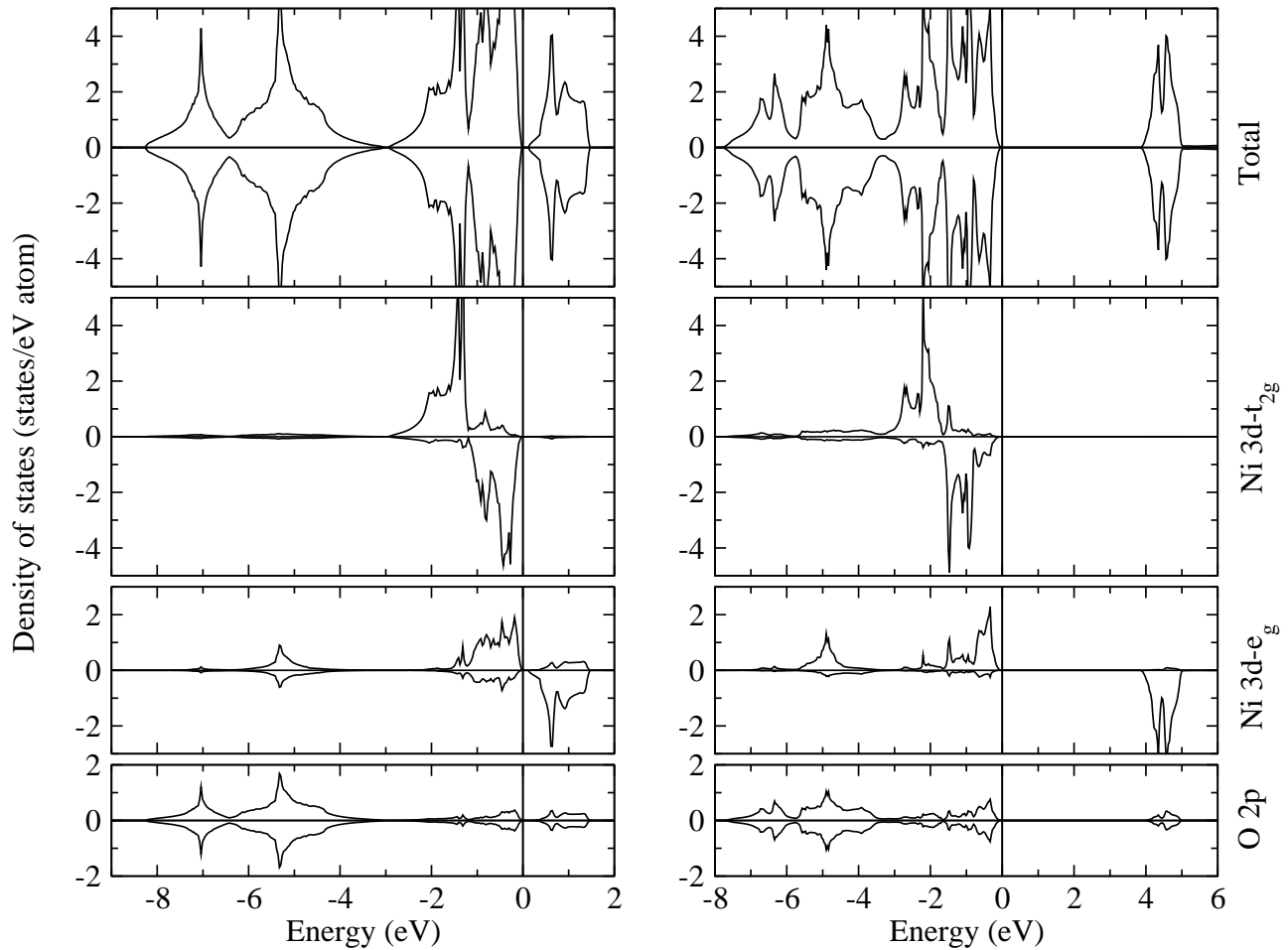}
  \caption{Total and partial densities of states for NiO obtained in LDA
  (left panel) and GTS (right panel) calculations.}
  \label{NiO-DOS}
\end{figure}
\end{widetext}

\begin{figure}
  \includegraphics[clip=true,width=0.45\textwidth]{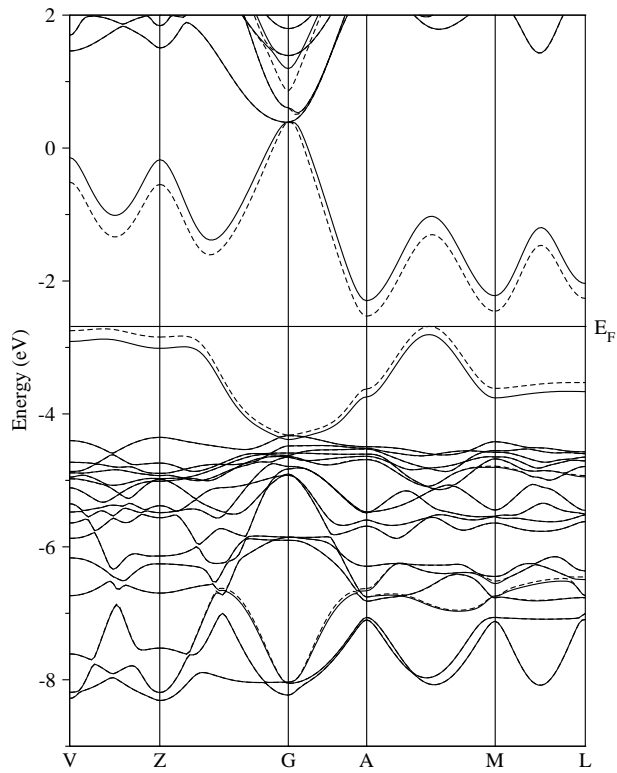}
  \caption{BaBiO$_3$ band structure. Dashed lines shows LDA results and solid lines
  correspond to GTS calculations}
  \label{BaBiO3-bands}
\end{figure}

LDA calculations for BaBiO$_3$ gave a negative or very small
positive values for energy gap \cite{licht,savrasov} while
spectroscopy measurements demonstrate a sizable value of
$\approx$0.5 eV\cite{BaBiO3_expt}. For our GTS calculations we
have chosen a minimal set of two Wannier functions calculated
using orbitals of Bi1-6s and Bi2-6s symmetry and bands located in
the energy window of $\pm$5 eV around Fermi energy. Constrained
LDA calculations for transition state gave the following
corrections to the energies of Wannier functions: $\delta
V_{Bi1-6s} = -0.19$ eV $\delta V_{Bi2-6s} = 0.43$ eV. The energy
band dispersions obtained in GTS calculations together with LDA
bands are shown on Fig.\ref{BaBiO3-bands}. Only two bands close to
Fermi energy were effected by GTS correction and energy gap value
has increased from 0.15 eV to a 0.48 eV value in good agreement
with experimental data (see Table \ref{result}).

\section{Conclusion}
\label{conclusion}

We developed a calculation scheme based on "transition state" idea
of Slater and Wannier functions set of one-electron states. This
method was applied to the four materials representing ionic
compounds (MgO), covalent bond semiconductors (Si), transition
metal oxide Mott insulators (NiO) and Peierls insulator BaBiO$_3$.
The results have shown a systematic improvement of energy gap
values. Not only excitation energies but also ground state
properties such as magnetic moment value for NiO can be
significantly improved comparing with LDA results. Encouraged by
the promising results reported in the present paper we plan to
apply this method to other materials where standard LDA approach
fails. Such calculations are in progress.

\section{Acknowledgments}
We are particulary indebted to Michael Korotin who contributed
significantly to this paper by his valuable criticism of our work.
This work was supported by Russian Foundation for Basic Research
under the grants RFFI-04-02-16096 and RFFI-03-02-39024, by the
joint UrO-SO Project N22, Netherlands Organization for Scientific
Research through NWO 047.016.005, programs of the Presidium of the
Russian Academy of Sciences (RAS) ``Quantum macrophysics'' and of
the Division of Physical Sciences of the RAS ``Strongly correlated
electrons in semiconductors, metals, superconductors and magnetic
materials''.

\begin{thebibliography}{99}
\bibitem{meaning} While beeing introduced as an auxiliary term to
derive "transition state" method equation
(\ref{const2}), this correction has effect of favoring
integer values of the occupancies (for $q_j=1$ and $q_j=0$ the
correction term has a minimum value) like in LDA+U
method\cite{ani97}.
\end{thebibliography}
\end{document}